\documentclass{www2006-submission}
\usepackage{times}
\def\more-auths{%
\end{tabular}
\begin{tabular}{c}}
\def\sharedaffiliation{%
\end{tabular}
\begin{tabular}{c}}

\begin{document}
\title{Decoding the structure of the WWW: facts versus sampling biases}
%
%

\numberofauthors{3}
%


\author{
%
\alignauthor M. \'{A}ngeles Serrano$^{1}$\\
       \email{mdserran@indiana.edu}
\alignauthor Ana Maguitman$^{1}$\\
       \email{anmaguit@cs.indiana.edu}
\alignauthor Mari\'{a}n Bogu\~{n}\'{a}$^{2}$\\
       \email{marian.boguna@ub.edu}
\vskip1pc  
\more-auths
\alignauthor Santo Fortunato$^{1,3}$ \\
      \email{santo@indiana.edu}
\alignauthor Alessandro Vespignani$^{1}$ \\
      \email{alexv@indiana.edu}
\vskip1pc  
\sharedaffiliation
      \affaddr{$^{1}$ School of Informatics, Indiana University}  \\
      \affaddr{Bloomington, IN 47406, USA}\\
      \affaddr{$^{2}$ Departament de F\'{i}sica Fonamental, Universitat de Barcelona}  \\
      \affaddr{08028 Barcelona, Spain}\\
      \affaddr{$^{3}$ Fakult\"at f\"ur Physik, Universit\"at Bielefeld} \\
      \affaddr{D-33501 Bielefeld, Germany}
}

\date{}

\maketitle

\begin{abstract}
The understanding of the immense and intricate topological structure
of the World Wide Web (WWW) is a major scientific and technological
challenge. This has been tackled recently by characterizing the
properties of its representative graphs in which vertices and
directed edges are identified with web-pages and hyperlinks,
respectively. Data gathered in large scale crawls have been analyzed
by several groups resulting in a general picture of the WWW that
encompasses many of the complex properties typical of rapidly
evolving
networks~\cite{Barabasi:1999,Broder:2000,Kumar:2000,Adamic:2001,Donato:2004}.
In this paper, we report a detailed statistical analysis of the
topological properties of four different WWW graphs obtained with
different crawlers. We find that, despite the very large size of the
samples, the statistical measures characterizing these graphs differ
quantitatively, and in some cases qualitatively, depending on the
domain analyzed and the crawl used for gathering the data. This
spurs the issue of the presence of sampling
biases~\cite{Henzinger:2000,Bar-Yossef:2000,Rusmevichientong:2001}
and structural differences of Web crawls that might induce
properties not representative of the actual global underlying graph.
In order to provide a more accurate characterization of the Web
graph and identify observables which are clearly discriminating with
respect to the sampling process, we study the behavior of
degree-degree correlation functions and the statistics of reciprocal
connections. The latter appears to enclose the relevant correlations
of the WWW graph and carry most of the topological information of
the Web. The analysis of this quantity is also of major interest in
relation to the navigability and searchability of the Web.
\end{abstract}

\category{H.4.m}{Information Systems}{Miscellaneous}
\category{G.3}{Mathematics and Computing}{Probability and
Statistics}

\terms{Measurement}

\keywords{Web graph structure, Web measurement, crawler biases,
statistical analysis}

\section{Introduction}

The World Wide Web (WWW) has grown at an unprecedented pace. While
it is not possible to provide a precise estimate of the WWW size in
terms of pages, a recent study~\cite{Gulli:2005}, which used Web
searches in 75 different languages, determined that there were over
11.5 billion Web pages in the publicly indexable
Web~\cite{Lawrence:1998,Lawrence:1999} at the end of January 2005.
Furthermore, the Web growth lacks any regulation and physical
constraint (contrary to what happens with the physical Internet
infrastructure~\cite{RomusVespasbook}), with new documents being
added or becoming obsolete very quickly.

A fundamental step in decoding and understanding the WWW
organization consists in the experimental studies of the WWW graph
structure in which vertices and directed edges are identified with
Web pages and hyperlinks, respectively.  These studies are based on
crawlers that explore the WWW connectivity by following the links on
each discovered page, thus reconstructing the topological properties
of the representative graph. Several studies based on those graphs
have been performed in order to reveal the large-scale topological
properties of the WWW. Distributions of in-degrees and out-degrees
have been found to exhibit heavy-tails and the macroscopic
architecture of connected components has made evident a rich
structural organization, i.e., the so-called bow-tie
structure~\cite{PLKumar,Barabasi:1999,Barabasi:2000,Broder:2000,Donato:2004}.
Reciprocal links and transitive relations regarding thematic
communities~\cite{Eckmann:2002} have attracted attention as well,
giving rise to a generally accepted picture of the topological
structure of the WWW.

While the importance of these studies is indisputable, the dynamical
nature of the Web and its huge size make very difficult the process
of compressing, ranking, indexing or mining the Web. Indeed, even
the largest scale Web crawlers cover only a small portion of the
publicly available information. In other words, it has been
impossible so far to achieve any complete unbiased large-scale
picture of the Web. On the other hand, the very large sizes of the
gathered data sets have led to the general belief that the
structural and statistical properties observed in the WWW graphs
were representative of the actual ones, thus leaving almost
untouched the study of possible sampling
biases~\cite{Henzinger:2000}. In this respect, on the one hand it is
crucial to understand clearly which is the exact information
provided by crawl engines, and, on the other hand, to explore to
which extent the Web properties we observe are not biased by the
specific characteristics of the crawls.

In this paper, we study four different data sets obtained in
different years with different crawls and for different domains of
the WWW. Our main contributions are:
\begin{itemize}
\item We provide a careful comparative analysis of the structural and statistical
topological properties of the different Web graphs, making evident
qualitative and quantitative differences across different samples.
We look at higher order statistical indicators characterizing single
and two-vertex correlations in order to provide a full account of
the connectivity pattern and structural ordering of the Web graph.
See Sections~\ref{SSP} and~\ref{SDDC}.

\item We identify a novel and crucial topological element, the reciprocal link, playing a key role
in the organization of the WWW and accounting for most of the
statistical correlations observed in Web graphs. Reciprocal
links~\cite{Reciprocity}, also referred in the literature as
bidirectional links~\cite{Boguna:2005} or
co-links~\cite{Eckmann:2002}, can allow us to clearly discriminate
among the statistical properties resulting from different crawls.
Furthermore, the inspection of the subgraphs of vertices
reciprocally connected provides interesting structural information
that might be crucial to assess how the underlying topology could
affect the functionality~\cite{Boguna:2005} of the Web and/or
processes running on it. Indeed, navigability and searchability are
intimately related to the functionality of the WWW, and those
properties strongly depend on the communication patterns among the
constituent sites of the network. See Section~\ref{SRL}.
\end{itemize}

\section{Related work}
The first empirical topological studies of the Web as a directed
graph focused on the measure of the directed degree distributions
$P(k_{in})$ and $P(k_{out})$, where the in/out-degree, $k_{in}$ or
$k_{out}$ respectively, is defined as the number of
incoming/outgoing links connecting a page to its neighbors. The work
by Kumar \textit{et al.}~\cite{PLKumar} on a big crawl of about 40M
nodes, and that by Barab\'{a}si and Albert~\cite{Barabasi:1999} on a
smaller set of over 0.3M nodes restricted to the domain of the
University of Notre Dame, suggested a scale-free nature for the WWW
with power-law behaviors both for the in- and out-degree
distributions.

Immediately after, a more complete investigation was published by
Broder \textit{et al.}~\cite{Broder:2000}. There, two sets from
AltaVista crawls were analyzed, corresponding to different months in
the same year 1999, May and October. The sets had over 200 million
pages and 1.5 billion links. The authors reported detailed
measurements on local and global properties of the Web graph which
covered, for instance, the degree distributions, corroborating
earlier observations, and also the presence and organization of
connected components, unfolding the so-called bow-tie structure of
the Web. One of the most intriguing conclusions there was that, from
the analysis of those two sets, the observed structure of the Web
was relatively insensitive to the particular large crawl used. In
addition, the connectivity structure of the Web was resilient to the
removal of a significant number of nodes.

Successively, further work~\cite{Donato:2004} along the same lines has
been performed over a large 2001 data set of 200M pages and about
1.4 billion edges made available by the WebBase project at Stanford (See next
section for references and a project description). In this work,
new measures were introduced along with the standard statistical
observables, and the obtained results were compared with the ones
presented in the work by Broder \textit{et al.}. One of the
reported differences is the deviation from the power-law behavior
of the out-degree distribution.

On the other hand, the question whether subsets of the Web display
the same characteristics as the Web at large has been discussed by a
number of authors. Dill \textit{et al.}~\cite{Dill:2001} found
self-similarity within thematically unified subgraphs extracted from
a single crawl of 60M pages gathered in October 2000. On the
contrary, the different components of the bow-tie decomposition have
been found to lack self-similarity in their inner structure when
compared to the whole graph~\cite{Donato:2005}.

\section{Data sets}
To gain some insight about how the crawling strategy affects
observations and on the existence of observable unbiased properties
we have analyzed and compared four sets of data corresponding to
different years, from 2001 to 2004, and different domains, general
and \textit{.uk} and \textit{.it} domains. The sets have been
gathered within two different projects: the WebBase project and the
WebGraph project, each using its own Web crawler, WebVac and
UbiCrawler respectively. {\bf The WebBase Project} is a World Wide
Web repository built as part of the Stanford Digital Libraries
Project by the Stanford University InfoLab
\footnote{http://www-db.stanford.edu/}. The Stanford WebBase
project\footnote{http://dbpubs.stanford.edu:8091/$\sim$testbed/doc2/WebBase/}~\cite{Hirai:2000}
is investigating various issues in crawling, storage, indexing, and
querying of large collections of Web pages. The project aims to
build the necessary infrastructure to facilitate the development and
testing of new algorithms for clustering, searching, mining, and
classification of Web content. The Stanford WebBase has been
collected by the spider WebVac~\cite{WebVac1,WebVac2} and makes
available a Web repository with access to general crawls, such as
the ones used in this research, or specific domain crawls
restricted, for instance, to universities or institutions. {\bf The
WebGraph Project}\footnote{http://webgraph.dsi.unimi.it/} is being
developed by the Laboratory for Web
Algorithmics\footnote{http://law.dsi.unimi.it/} (LAW) at the
University of Milano and analyzes data obtained by its own crawler,
UbiCrawler\footnote{http://ubi.iit.cnr.it/projects/ubicrawler/}~\cite{Boldi:2004b},
designed to achieve high scalability and to be tolerant to failures.

The above projects provide several data sets publicly available to
researchers. We analyze four samples ranging from 2001 to 2004. The
WebBase general crawl of 2001 (WBGC01) and the WebBase general crawl
of 2003 (WBGC03)\footnote{ftp://db.stanford.edu/pub/webbase/} have
been collected by the WebBase project in a general crawl using the
WebVac spider. The remaining two sets collected by the UbiCrawler
project, the WebGraph \textit{.uk} domain of 2002
(WGUK02)\footnote{http://webdata.iit.cnr.it/united\b{
}kingdom-2002/} and WebGraph \textit{.it} domain of 2004
(WGIT04)\footnote{http://webdata.iit.cnr.it/italy-2004/}, are
restricted to the domains \textit{.uk} and \textit{.it},
respectively. Note that the two domain crawls present an interesting
difference. While pages in the \textit{.uk} domain have higher
probability to point to pages outside the domain, due to English
being the official language in other influential countries, such as
the USA, and to the widespread use of English, the links in the
Italian \textit{.it} domain may be much more endogenous, which could
potentially have a high effect on the Web description derived from
the data.

We have cleaned the four sets by disregarding multiple links between
the same pages and self-connections. In Table \ref{tab:features} we
present a summary of the size in vertices and directed edges of the
four sets analyzed in this paper.
\begin{table}
\centering \caption{Number of nodes and edges of the networks
considered, after extracting multiple links and self-connections.}
\label{tab:features}\vspace{0.2cm}{\scriptsize
\begin{tabular}{|c||c||c||c||c|}
\hline &&&&\\[-5.5pt]
Data set & WBGC01 & WGUK02 & WBGC03 & WGIT04 \\
\hline \hline &&&&\\[-5.5pt]\# nodes & 80571247 &  18520486 & 49296313 & 41291594\\[1pt]
\hline &&&&\\[-5.5pt] \# links & 752527660 & 292243663  & 1185396953 & 1135718909\\[1pt]
\hline
\end{tabular}}
\end{table}

All the following measures have been carried out using Matlab
code.~\footnote{Available upon request.}

\section{Structural properties}
\label{SSP} Data gathered in large scale
crawls~\cite{PLKumar,Barabasi:1999,Barabasi:2000,Broder:2000,Eckmann:2002,Donato:2004}
have uncovered the presence of a complex architecture underlying the
structure of the Web graph. A widespread feature is the small-world
property. Despite its huge size, the average number of URL links
that must be followed to navigate from one document to the other,
technically the average shortest path length, seems to be very small
as compared to the value for a regular lattice of comparable size,
and it seems to grow with the system size very slowly at a
logarithmic pace~\cite{SmW,Broder:2000}. Another important result is
that the WWW exhibits a power-law relationship between the frequency
of vertices and their degree, defined as the number of directed
edges linking each vertex to its neighbors. This last feature is the
signature of a very complex and heterogeneous topology with
statistical fluctuations extending over many length
scales~\cite{SmW,Barabasi:1999,PLKumar}. Finally, a fascinating
macroscopic description of the Web has been provided by the study of
the connected components, taking into account the directed nature of
the Web graph~\cite{Broder:2000}. In the following, we perform a
careful comparative analysis of the four Web crawls described in the
previous section. This will allow us to critically examine the
stability of the various results as a function of the crawl and
discuss which properties appear to be genuine features of the global
Web graph.

\subsection{Sizes of connected components}
\label{sec_cc} 
\begin{figure}[t]
\centering \epsfig{file=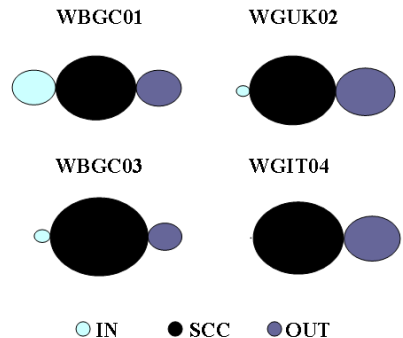,width=6cm} \caption{Graphical
representation of the sizes of the global components reported in
Table \ref{tab:componentsO}. The area of each component is
proportional to its actual size, so that the relative sizes of the
components in the figure account for the actual relative sizes of
the Web graphs.} \label{GSCC}
\end{figure}

The directed nature of the Web brings out a complex
structure of connected components~\cite{RomusVespasbook,Mendesbook}
that has been captured in the famous bow-tie architecture
highlighted in the study presented in~\cite{Broder:2000}. If we
disregard the directedness of links, the weakly connected component
of the graph is made by all pages belonging to the giant component
of the corresponding undirected graph. The undirected component
becomes internally structured when the directed nature of the
connections is considered. The most important of these new internal
components is called the strongly connected component (SCC), which
includes all pages  mutually connected by a directed path. The other
two relevant components are the in-component (IN) and the
out-component (OUT). The first is formed by the vertices from which
it is possible to reach the SCC by means of a directed path. The
second refers to the set of vertices that can be reached from the
SCC by means of a directed path. Finally, other secondary structures
can also be present, such as tendrils, which contain pages that
cannot reach the SCC and cannot be reached from it, or tubes which
can directly connect the IN and OUT components without crossing the
SCC. This complex composition is usually called the bow-tie
structure because of the typical shape assumed by the figure
sketching the relative size of each component (see Fig.~\ref{GSCC}).
It is clear that such a component structure is extremely relevant in
the discussion of the functionalities of the Web. For instance, the
relative sizes of the SCC and the IN and OUT components give us
information about the probabilities of returning to an original page
after exploration, or the size of the accessible Web once a starting
page has been selected. The size of the SCC is of particular
importance, since it constitutes the subset of reversible and
complete access navigability. When one starts to surf the Web from
the IN component, it is very likely that after a while one ends up
in the SCC, and maybe eventually in the OUT component, but can never
go back to the original point. Once in the OUT component, one can
never go back to the other main components. But within the SCC, all
nodes are reachable and can be revisited.
\begin{table}
\centering \caption{Sizes of the SCC, IN and OUT components and
their sum MAIN$=$SCC$+$IN$+$OUT. Notice that MAIN does not contain
either tendrils or tubes, so that it differs from the weakly
connected component. Values are shown as a percentage of the total
number of nodes.} \label{tab:componentsO}\vspace{0.2cm}{\scriptsize
\begin{tabular}{|c||c||c||c||c|}
\hline &&&&\\[-5.5pt] Data set & WBGC01 & WGUK02 & WBGC03 & WGIT04 \\
\hline \hline &&&&\\[-5.5pt] IN & 17.24 & 1.69  & 2.28 &0.03\\[1pt]
\hline &&&&\\[-5.5pt] SCC &56.46&65.28 &85.87 &72.30\\[1pt]
\hline &&&&\\[-5.5pt] OUT&17.94 &31.88 &11.26 &27.64 \\[1pt]
\hline \hline &&&&\\[-5.5pt] MAIN&91.64 &98.85 &99.41 &99.98 \\[1pt]
\hline
\end{tabular}}
\end{table}

We summarize the values for the sizes of the components of the four
data sets in Table \ref{tab:componentsO}. The figures for the domain
crawls are in agreement to those reported in~\cite{Donato:2005},
where the same \textit{.uk} and \textit{.it} sets were also
examined. The analysis of the four data sets considered in the
present study shows a noticeable variability of the basic component
structure of the resulting graph. In particular, the IN component is
the most unstable feature that ranges from accounting for about 20\%
of the total structure (WBGC01) to the case in which it is
practically absent (WGIT04). This variability could be likely
ascribed to the different crawling strategies and the fact that each
of those may use different starting points. Moreover, crawlers
perform a directed exploration in the sense that they follow
outgoing hyperlinks to reach pointed pages, but cannot navigate
backwards using incoming hyperlinks. This implies that the
exploration of the IN component is strongly biased by the initial
conditions used to start the crawl. Variations are however not
limited to the IN component. Also the relative sizes of the SCC and
the OUT component vary from sample to sample, even by a factor close
to three in the case of the OUT component. Finally, notice that the
sizes of the IN and OUT components of the WBGC01 set are quite
symmetric, as was also found in~\cite{Broder:2000}, where the values
reported for the sizes of the IN, SCC and OUT of components of the
AltaVista crawl were $21.3\%, 27.7\%, 21.2\%$ respectively. In
summary, it is evident from this analysis that the structure of Web
graphs is strongly dependent on the crawler strategies.

\subsection{Degree distributions}
\label{sec_degreedistributions} 
\begin{figure}[t]\hspace{-0.3cm}
\centering \epsfig{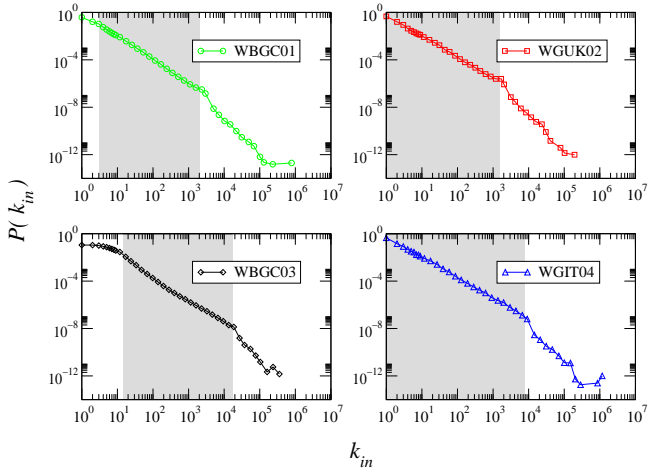}
\caption{Distributions of incoming links. In the shadowed regions
all the functions decay as a power-law with exponents given in
Table~\ref{tab:gammasin}.} \label{fig:in}
\end{figure}

\begin{figure}
\centering \epsfig{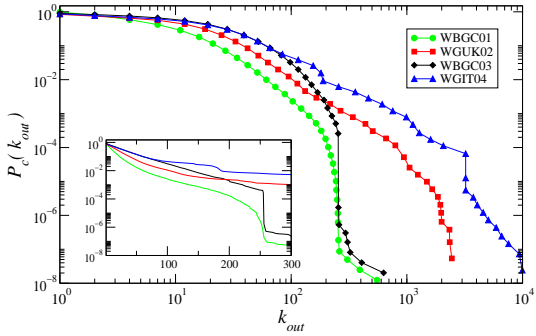}
\caption{Distributions of outgoing links. For visualization
purposes, we use cumulative distributions defined as
$P_c(k_{out})=\sum_{k'_{out}\geq k_{out}}P(k'_{out})$. The inset
shows the same curves in a linear-log scale.} \label{fig:out}
\end{figure}

A major interesting feature found in
Web graphs is the presence of a highly heterogeneous topology, with
degree distributions characterized by wide variability and heavy
tails~\cite{SmW,Barabasi:1999,PLKumar}. The degree distribution
$P(k)$ for undirected networks is defined as the probability that a
node is connected to $k$ other nodes. For directed networks, this
function splits in two separate functions, the in-degree
distribution $P(k_{in})$ and the out-degree distribution
$P(k_{out})$, which are measured separately as the probabilities of
having $k_{in}$ incoming links and $k_{out}$ outgoing links,
respectively. In Figs.~\ref{fig:in} and \ref{fig:out} we report the
behavior of the in-degree and out-degree distributions.
These distributions, as for most real world networks, are found to
be very different from the degree distribution of a random graph or
an ordered lattice. They are both skewed and spanning several orders
of magnitude in degree values. The in-degree distribution exhibits a
heavy-tailed form approximated by a power-law behavior
$P(k_{in})\sim k_{in}^{-\gamma_{in}}$, generally spanning over 3 to
4 orders of magnitude. In Figure~\ref{fig:in}, we show the region
considered in the evaluation of the exponent obtained by a maximum
likelihood algorithm for discrete distributions. The in-degree
distributions also exhibit a noisy tail that cannot be well fitted
with a specific analytic form. Yet it strengthens the evidence for
the heavy-tailed character of $P(k_{in})$.
\begin{table}
\centering \caption{Main statistical properties of the analyzed
sets: average degree $\langle k\rangle$, maximum degree $k_{max}$,
standard deviation $\sigma$, heterogeneity parameter $\kappa$, and
maximum likelihood estimate of the exponent of the power-law
in-degree distribution $\gamma_{in}$ (precision error $\pm 0.1$).
All values are provided for in- and out-degrees and for the four
data sets. The symbol $\infty$ for $\gamma_{out}$ means that the
out-degree distributions decay faster than a power-law.}
\vspace{0.2cm}\label{tab:gammasin}\vspace{0.2cm} {\scriptsize
\begin{tabular}{|c||c||c||c||c|}
\hline &&&&\\[-5.5pt] Data set & WBGC01 & WGUK02& WBGC03 & WGIT04 \\
\hline \hline &&&&\\[-5.5pt] $\langle k_{in}\rangle$ &9.3 &15.8 &24.1 &27.5\\[1pt]
 &&&&\\[-5.5pt]  $k_{in}^{max}$ &788632&194942 &378875 &1326744\\[1pt]
 &&&&\\[-5.5pt] $\sigma_{in}$&200.2 &143.3 &421.6 &881.4 \\[1pt]
 &&&&\\[-5.5pt]  $\kappa_{in}$ &4298.6 &1317.5 &7414.9&28269.9\\[1pt]
 &&&&\\[-5.5pt] $\gamma_{in}$&1.9  &1.7 &2.2&1.6 \\[1pt]
\hline \hline &&&&\\[-5.5pt] & WBGC01 & WGUK02 & WBGC03 & WGIT04 \\
\hline \hline &&&&\\[-5.5pt] $\langle k_{out}\rangle$ &9.3  &15.8&24.1&27.5\\[1pt]
 &&&&\\[-5.5pt]  $k_{out}^{max}$ &552 &2449 &629&9964\\[1pt]
 &&&&\\[-5.5pt] $\sigma_{out}$&13.1  &27.4 &29.5&67.1\\[1pt]
 &&&&\\[-5.5pt]  $\kappa_{out}$ &27.7 &63.4&60.3 &191.0\\[1pt]
 &&&&\\[-5.5pt] $\gamma_{out}$&$\infty$&$\infty$ & $\infty$&$\infty$ \\[1pt]
\hline
\end{tabular}}
\end{table}

A different situation is faced in the case of the out-degree
distribution $P(k_{out})$. In this case, a clear exponential cut-off
is observed and the range of degree values is 2 to 4 orders of
magnitude smaller than what found for the in-degree distribution.
The origin of the cut-off can be explained by the different nature
of the in-degree and out-degree evolution. The in-degree of a vertex
is the sum of all the hyperlinks incoming from all the Web pages in
the WWW. In principle, thus, there is no limit to the number of
incoming hyperlinks, that is determined only by the popularity of
the Web page itself. On the contrary, the out-degree is determined
by the number of hyperlinks present in the page, which are
controlled by Web administrators. For evident reasons (clarity,
handling, data storage) it is very unlikely to find an excessively
large number of hyperlinks in a given page. This represents a sort
of finite capacity~\cite{Amaral:2002} for the formation of outgoing
hyperlinks that might naturally lead to a finite cut-off in the
out-degree distribution.

The heavy-tailed behavior of the in-degree distribution implies that
there is a statistically significant probability that a vertex has a
very large number of connections compared to the average degree
$\langle k_{in} \rangle$. In addition, the extremely large value of
$\langle k_{in}^2 \rangle$, and therefore of the variance
$\sigma^2=\langle k_{in}^2 \rangle- \langle k_{in} \rangle^2$ is
signalling the extreme heterogeneity of the connectivity pattern,
since it implies that statistical fluctuations are virtually
unbounded, and tells us that the average degree is not the typical
degree value in the system, i.e., we have scale-free distributions.
The heavy-tailed nature of the degree distribution has also
important consequences in the dynamics of processes taking place on
top of these networks. Indeed, recent studies about network
resilience to removal of vertices~\cite{Havlin:2000} and
spreading~\cite{Vespignani:2001} have shown that the relevant
parameter for these phenomena is the ratio between the first two
moments of the degree distribution $\kappa=\langle
k^2\rangle/\langle k \rangle$. If $\kappa\gg1$ the network manifests
some properties that are not observed for networks with
exponentially decaying degree distributions. In the case of directed
networks, this heterogeneity parameter has to be defined separately
for in- and out-degrees as $\kappa_{in}=\langle
k_{in}^2\rangle/\langle k_{in} \rangle$ and $\kappa_{out}=\langle
k_{out}^2\rangle/\langle k_{out} \rangle$,~\footnote{Notice that for
any directed graph $\langle k_{in} \rangle=\langle k_{out}
\rangle$.} since it could happen that the network is heterogeneous
with respect to one of the degrees but not to the
other.~\footnote{In addition, a third parameter can be defined which
accounts for the effect of the crossed one point correlations
$\kappa_{in,out}=\langle k_{in} k_{out} \rangle/\langle k_{in}
\rangle$.} In Table~\ref{tab:gammasin}, we provide these values for
the empirical graphs along with a summary of the numerical
properties of the probability distributions analyzed so far. The
heavy-tailed behavior is especially evident when comparing the
heterogeneity parameters $\kappa$ and their wide range variations. A
marked difference is observed for the out-degree distributions where
the variance and heterogeneity parameters are indicating a limited
variability of the function $P(k_{out})$. From the exponents
reported for the in-degree distribution, it results evident that the
fittings to a power-law form can yield slightly different results,
depending on the data set analyzed. These variations could signal a
slightly different structure of the Web graph depending on the
domain crawled or the eventual presence of statistical biases due to
the crawling strategy. It is interesting to notice that a similar
variability is encountered in studies of the power-law behavior of
Web samples restricted to specific thematic
groups~\cite{Pennock:2002}. Another oddity that has to be signalled
is the fact that the general crawls WBGC01 and WBGC03 exhibit a much
smaller cut-off of the out-degree distribution than observed in the
two domain crawls. This is somehow counterintuitive given the larger
sizes of the general crawls. This might hint to the presence of a
bias in the way hyperlinks are explored by different crawlers, again
purporting evidence for the presence of sampling  biases that affect
the observed statistical properties of Web graphs.

\section{Degree correlations}
\label{SDDC} As an initial discriminant of structural ordering, the
attention has been focused on the networks' degree distribution.
This function is, however, only one of the many statistics
characterizing the structural and hierarchical ordering of a
network; a full account of the connectivity pattern calls for the
detailed study of degree correlations. Along these lines, for
instance, it is possible to provide a quantitative study of the
mixing properties of networks through opportune projection of the
degree-degree joint probability distribution. This allows the
distinction between assortative networks, in which large degree
nodes preferentially attach to large degree nodes, and
disassortative networks, showing the opposite
tendency~\cite{AssortativeNewman}. These structural properties are
the signature of specific ordering principles.

\subsection{Single vertex degree correlations}
\begin{figure}
\centering \psfig{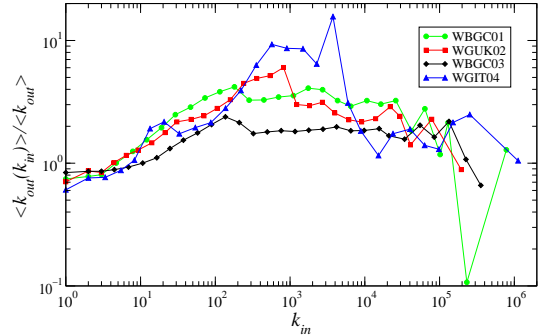} \caption{Normalized
average out-degree as a function of the in-degree for the four
different data sets.} \label{onecorr}
\end{figure}

First, we examine local one-point degree correlations for individual
nodes, in order to understand if there is a relation between the
number of incoming and outgoing links in single pages. Since most of
the analyzed degree distributions are heavy-tailed, fluctuations are
extremely large so that the linear correlation coefficient is not
well defined for those cases. Instead, we provide the crossed
one-point correlations, $\langle k_{in} k_{out}\rangle$, normalized
by the corresponding uncorrelated value, $\langle k_{in}\rangle
\langle k_{out}\rangle$. We also report the function
\begin{equation}
\langle k_{out}(k_{in})\rangle=\frac{1}{N_{k_{in}}}\sum_{i \in
\Upsilon(k_{in})} k_{out,i},
\end{equation}
which measures the average out-degree of nodes as a function of
their in-degree. $N_{k_{in}}$ stands for the number of nodes with
in-degree $k_{in}$ and $k_{out,i}$ is the out-degree of node $i$.
The notation $i \in \Upsilon(k_{in})$ indicates that the summation
has to be performed over the set of nodes of degree $k_{in}$,
denoted by $\Upsilon(k_{in})$. The results can be found in Table
\ref{tab:correlations} and in Fig.~\ref{onecorr}.
\begin{table}
\centering \caption{Crossed in-degree out-degree correlations for
individual nodes, normalized by the uncorrelated values.}
\label{tab:correlations}\vspace{0.2cm}{\scriptsize
\begin{tabular}{|c||c||c||c||c|}
\hline &&&&\\[-5.5pt] Data set & WBGC01 & WGUK02 & WBGC03 & WGIT04 \\
\hline \hline &&&&\\[-5.5pt]
$\frac{\langle k_{in} k_{out}\rangle}{\langle k_{in}\rangle \langle k_{out}\rangle}$ &2.8 & 3.1 &1.6 &5.6\\[3pt]
\hline
\end{tabular}}
\end{table}

A significant positive correlation between the in-degrees and the
out-degrees of single nodes is found for all the sets. That means
that more popular pages tend to point to a higher number of other
pages. This positive correlation is found to be true for a range of
in-degrees that spans from $k_{in}=1$ to $k_{in}=10^2\sim10^3$,
depending on the specific set. Beyond this point no noticeable
correlation is observed, see Fig. \ref{onecorr}. The set for the
Italian domain is more noisy, but this pattern appears to be
independent of the crawl used to gather the data and, thus, it seems
to be a genuine feature of the Web.

\subsection{Two-vertex degree correlations}
\label{degree-degree} Another important source of information about
the network structural organization lies in the correlations of the
degrees of neighboring vertices. These correlations can be probed in
undirected networks by inspecting the average degree of nearest
neighbors of a vertex $i$, where nearest neighbors refers to the set
of vertices at a hop distance equal to 1,
\begin{equation}
  \overline{k_{nn,i}}= \frac{1}{k_i}\sum_{j\epsilon \nu(i)}k_j.
   \label{eq_knni}
\end{equation}
The sum runs over the nearest neighbor vertices of each vertex $i$,
gathered in the set $\nu(i)$. From this quantity, a convenient
measure is obtained by averaging over degree classes to obtain the
average degree of the nearest neighbors for vertices of degree $k$,
defined as~\cite{InternetEmpiricalVespas}
\begin{equation}
  \overline{k_{nn}}(k) = \frac{1}{N_k}\sum_{i \in \Upsilon(k)}
  \overline{k_{nn,i}}=\sum_{k'}k'P(k'|k),
  \label{eq_knn}
\end{equation}
where $N_k$ is the number of nodes with degree $k$, the notation $i
\in \Upsilon(k)$ indicates that the summation has to be performed
over the set of nodes of degree $k$, denoted by $\Upsilon(k)$, and
$P(k'|k)$ quantifies the conditional probability that a vertex with
degree $k$ is connected to a vertex with degree $k'$. This measure
provides a sharp proof of the presence or absence of correlations.
In the case of uncorrelated networks, the degrees of connected
vertices are independent random quantities, so that $P(k'|k)$ is
only a function of $k'$. In this case, $\overline{k_{nn}}(k)$ does
not depend on $k$ and equals $\kappa=\langle k^2\rangle/\langle
k\rangle$. Therefore, a function $\overline{k_{nn}}(k)$ showing any
explicit dependence on $k$ signals the presence of degree
correlations in the system. Real networks usually tend to display
one of two different patterns~\cite{AssortativeNewman}. Assortative
networks exhibit $\overline{k_{nn}}(k)$ functions increasing with
$k$, which denotes that vertices are preferentially connected to
other vertices with similar degree. Examples of assortative behavior
are typically found in many social structures. On the other hand,
disassortative networks exhibit  $\overline{k_{nn}}(k)$ functions
decreasing with $k$, which denotes that vertices are preferentially
connected to other vertices with very different degree. Examples of
disassortative behavior are typically found in several technological
networks, as well as in communication and biological networks.

\begin{figure}[t]
\centering \epsfig{file=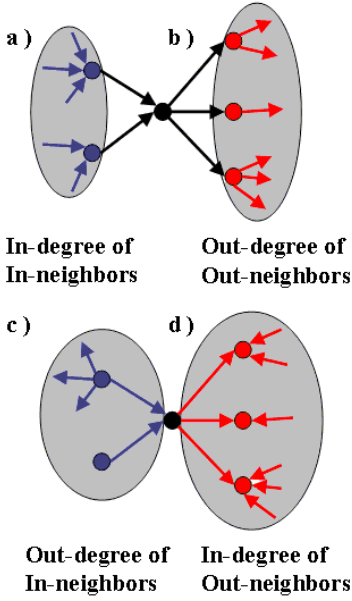,width=5.5cm} \caption{Graphical
sketch illustrating the degree-degree correlation functions defined
in section~\ref {degree-degree}. We focus on a single node --the
central node in the figures-- with in-degree $k_{in}=2$ and
out-degree $k_{out}=3$. In a) the average in-degree of its
in-neighbors is computed taking into account the incoming arrows
inside the grey area. The function $\overline{k_{in,nn}}(k_{in})$ is
then the average of this quantity over all nodes with the same
in-degree. The rest of the functions are defined in a similar way,
as highlighted in b), c), and d).} \label{knnsd}
\end{figure}
\begin{figure}
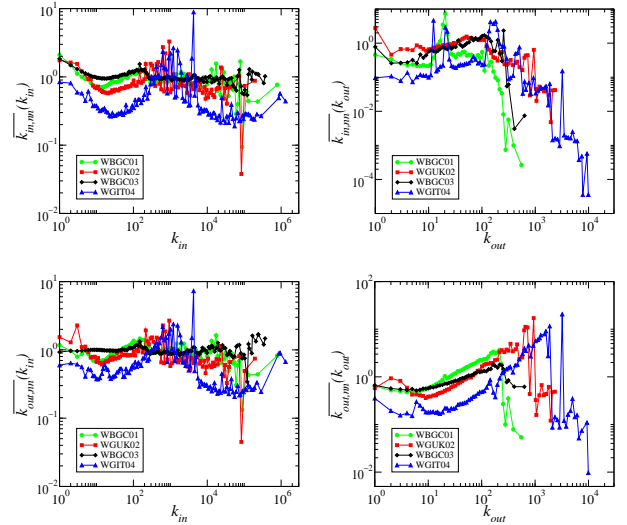

\centering
$\begin{array} {c@{\hspace{0.15in}}c}
  \psfig{file=knninin.eps,width=1.5in} &
  \psfig{file=knninout.eps,width=1.5in} \\[0.2cm]
  \psfig{file=knnoutin.eps,width=1.5in} &
  \psfig{file=knnoutout.eps,width=1.5in}
\end{array}$
\caption{Degree-degree correlations for the four different data
sets. Explicit expressions for the quantitative definition of these
functions can be found in Appendix~\ref{APPA}.} \label{knn1}
\end{figure}

In the case of the WWW, the study of the degree-degree correlation
functions is naturally affected by the directed nature of the graph.
In~\cite{WWWVespas}, a set of directed degree-degree correlation
functions was defined considering that, in this case, the neighbors
can be restricted to those connected by a certain type of directed
link, either incoming or outgoing. For the WWW, we study the most
significant distributions, taking into account that we can partition
the neighborhood of each single node $i$ into neighboring nodes
connected to it by incoming links and neighboring nodes connected to
it by outgoing links. A first correlation indicator,
$\overline{k_{in,nn}}(k_{in})$, is defined as the normalized average
in-degree of the neighbors of nodes of in-degree $k_{in}$, when
those neighboring nodes are found following incoming links of the
original node, see Fig.~\ref{knnsd} a). If we measure the popularity
of Web pages in terms of the number of pages pointing to them, this
function quantifies the average popularity of pages pointing to
pages with a certain popularity. The exact definition is given in
Appendix~\ref{APPA} along with the expression for the normalization
factor. The rest of the correlation functions,
$\overline{k_{out,nn}}(k_{in})$, $\overline{k_{out,nn}}(k_{out})$,
$\overline{k_{in,nn}}(k_{out})$ can be defined in an analogous
manner. Each plot in Fig.~\ref{knn1} shows these correlation
functions for the four data sets analyzed in this paper.
Remarkably, only one of the functions shows an increasing pattern
denoting the presence of assortative correlations for the four data
sets. The average out-degree of neighbors of nodes of high
out-degree is also high, so that the average number of references is
high in pages pointed by pages with a high number of references. In
all other cases, very mild or a complete lack of correlation is
observed. This is somehow surprising since, from the observed
similarities in the correlation patterns, one cannot infer the
differences in the structural properties observed in
Sec.~\ref{sec_cc} for the different Web graphs.

\section{The role of reciprocal links}
\label{SRL} While a directed network, the Web has many pages
pointing to each other. A couple of pages pointing to each other
corresponds to the presence of a reciprocal link that can be
considered as undirected. These reciprocal connections play an
important role and in this section we introduce and investigate
reciprocal links as crucial elements in the understanding of the
WWW. To this end, we will differentiate into incoming, outgoing, and
reciprocal links, where incoming and outgoing links do not include
the ones taking part in reciprocal connections and are referred to
as non-reciprocal. This allows us to consider reciprocal and
non-reciprocal connections as separate and well-defined independent
entities and provides a statistical analysis able to capture
additional information of the Web structure and the sampling biases
eventually observed in different data sets.

\subsection{Degree distributions}
\begin{figure} \centering
\epsfig{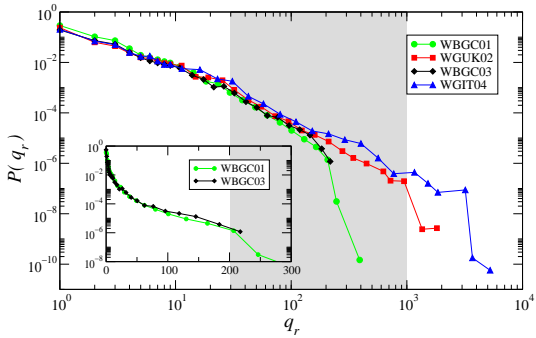} \caption{Probability
distributions of reciprocal links. The inset shows the distributions
for the two general crawls in a linear-log scale.} \label{fig:b}
\end{figure}

For the sake of notation, in the following we will identify the
non-reciprocal in-degree and out-degree of a given vertex $i$ with
$q_{in,i}$ and $q_{out,i}$, respectively. Analogously, the
reciprocal degree (r-degree) $q_{r,i}$ indicates the number of
reciprocal connections to neighboring vertices. While the degree
distributions of non-reciprocal links are extremely similar to those
obtained for the global in and out-degree, the reciprocal degree
distribution appears to exhibit a striking different behavior
depending on the crawl examined. In particular, general crawls show
a distribution $P(q_r)$ with an exponentially fast decaying
behavior, while the domain crawls have a heavy-tailed distribution
varying over three orders of magnitude (see Fig.~\ref{fig:b}). In
Table~\ref{tab:gammasb}, we summarize the main properties of
$P(q_r)$ for the various data sets. Also from the values shown there
one can easily see the mild fluctuations and heterogeneity expressed
by the general crawl data sets.
\begin{table}
\centering \caption{Main statistical properties of the reciprocal
subgraphs: average degree $\langle q_r\rangle$, maximum degree
$q_r^{max}$, standard deviation $\sigma_r$, heterogeneity parameter
$\kappa_r$, and maximum likelihood estimate of the exponent of the
power-law in-degree distribution $\gamma_r$ (precision error $\pm
0.1$). The symbol $\infty$ means that the distribution decays faster
than a power-law.} \vspace{0.2cm}\label{tab:gammasb}\vspace{0.2cm}
{\scriptsize
\begin{tabular}{|c||c||c||c||c|}
\hline &&&&\\[-5.5pt] Data set & WBGC01 & WGUK02& WBGC03 &WGIT04 \\
\hline \hline &&&&\\[-5.5pt] $\langle q_{r}\rangle$ &2.7 &3.3 &2.4&5.2\\[1pt]
 &&&&\\[-5.5pt]  $q_{r}^{max}$ &391&1997 &253 &6164\\[1pt]
 &&&&\\[-5.5pt] $\sigma_{r}$&7.2 &16.2 &8.1&42.7 \\[1pt]
 &&&&\\[-5.5pt]  $\kappa_{r}$ &21.9 &82.7 &30.0&352.6\\[1pt]
 &&&&\\[-5.5pt] $\gamma_{r}$&$\infty$  &2.6 &$\infty$&2.6 \\[1pt]
\hline
\end{tabular}}
\end{table}
The evident differences in the reciprocal degree distributions match
the dissimilar component structure observed in general and domain
crawls. On the other hand, the origin of the two different
statistical behaviors does not find a clear explanation and deserves
further investigation. In particular, it is not possible to find an
easy explanation either in the crawling strategies or in the
eventual features of Web specific domains. Finally, once again we
have to emphasize the odd finding of general crawls showing
reciprocal degree distribution cut-offs much smaller than those
observed for domain crawls.

\subsection{One-point degree correlations}
\begin{figure}[t]
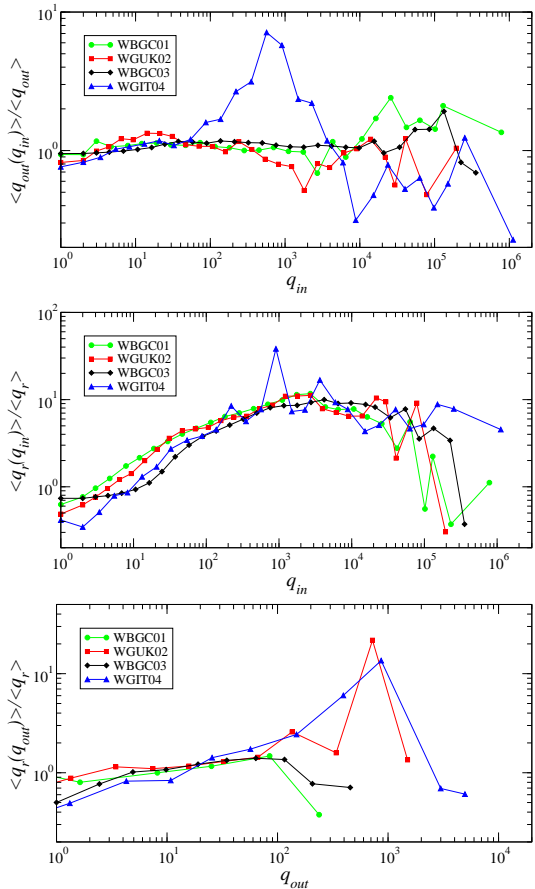

\centering $\begin{array} {c}
  \psfig{file=koutckinc.eps,width=7cm} \\
  \psfig{file=kbkinc.eps,width=7cm} \\
  \psfig{file=kbkoutc.eps,width=7cm}
\end{array}$
\caption{One node correlations for the four different data sets. The
functions shown are the normalized average non-reciprocal out-degree
as a function of the non-reciprocal in-degree, and the normalized
average r-degree as a function of the non-reciprocal in- and
out-degrees.} \label{onecorrclean}
\end{figure}

\begin{table}
\centering \caption{Crossed non-reciprocal in-degree, out-degree,
and r-degree correlations for individual nodes.}
\label{tab:correlationsclean}\vspace{0.2cm}{\scriptsize
\begin{tabular}{|c||c||c||c||c|}
\hline &&&&\\[-5.5pt] Data set & WBGC01 & WGUK02 & WBGC03 & WGIT04 \\
\hline \hline &&&&\\[-5.5pt] $\frac{\langle q_{in} q_{out}\rangle}{\langle q_{in}\rangle \langle q_{out}\rangle}$ &1.0 & 0.9 &1.1 &2.0\\[3pt]
\hline &&&&\\[-5.5pt] $\frac{\langle q_{in} q_{r}\rangle}{\langle q_{in}\rangle \langle q_{r}\rangle}$ &6.7 &7.4  & 6.0&9.9\\[3pt]
\hline &&&&\\[-5.5pt] $\frac{\langle q_{out} q_{r}\rangle}{\langle q_{out}\rangle \langle q_{r}\rangle}$ &1.1 &  1.4&1.3 &2.4\\[3pt]
\hline
\end{tabular}}
\end{table}
The distinction between reciprocal and non-reciprocal links induces
a higher complexity even at the most local level. In this case, each
node is characterized by three different quantities. Consequently,
we need to introduce three correlation measures, i.e., the average
non-reciprocal out-degree as a function of the non-reciprocal
in-degree, $\langle q_{out}(q_{in})\rangle$, and the average
r-degree as a function of the number of non-reciprocal incoming and
outgoing links, $\langle q_{r}(q_{in}) \rangle$ and $\langle
q_{r}(q_{out})\rangle$, respectively (see Fig. \ref{onecorrclean}).
A surprising result is that, in this case, there is no clear
correlation between non-reciprocal in- and out- degrees but there is
a positive correlation between reciprocal and non-reciprocal
in-degrees. So, the positive correlation previously observed between
in- and out-degrees is just a consequence of this new correlation.

\subsection{Degree-degree correlations}
The two vertices correlation analysis presented in section 5.2 can
be repeated for the non-reciprocal and reciprocal decomposition of
the network. Now, we have to differentiate reciprocal links and
segregate the neighborhood of each single node $i$ into neighboring
nodes connected to it by non-reciprocal incoming links, neighboring
nodes connected to it by non-reciprocal outgoing links, and
neighboring nodes connected to it by reciprocal links. The
degree-degree correlation functions corresponding to the first two
cases give similar results to the ones presented in the previous
section and do not signal the presence of any relevant correlation
pattern (not plotted).

\begin{figure}[t]
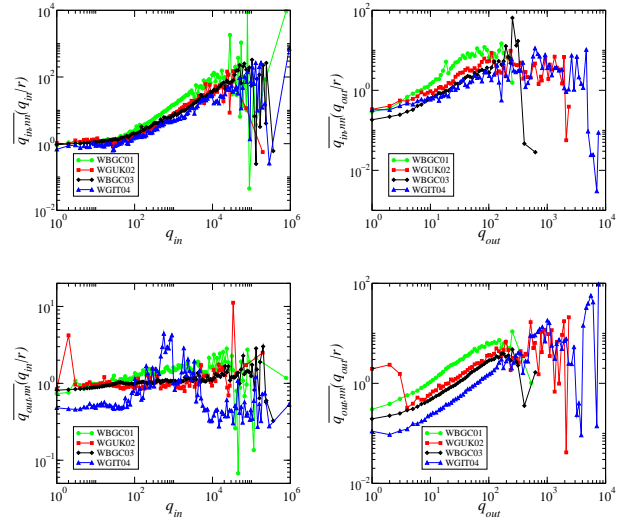

\centering $\begin{array} {c@{\hspace{0.15in}}c}
  \psfig{file=knnininbidi.eps,width=1.5in} &
  \psfig{file=knninoutbidi.eps,width=1.5in} \\[0.3cm]
  \psfig{file=knnoutinbidi.eps,width=1.5in} &
  \psfig{file=knnoutoutbidi.eps,width=1.5in}
\end{array}$
\caption{Non reciprocal degree-degree correlations for the four
different data sets.} \label{knnnr}
\end{figure}

A very different picture is obtained when we measure correlations
following reciprocal connections. A strong positive correlation is
observed between the in-degrees of nodes connected by reciprocal
links. This is clearly visible in the upper left plot of
Fig.~\ref{knnnr}, which shows the normalized average non-reciprocal
in-degree of the neighbors of nodes of non-reciprocal in-degree
$q_{in}$, when the neighbors are found following reciprocal links,
$\overline{q_{in,nn}}(q_{in}|r)$. This function shows a clear
increase of two orders of magnitude as a function of $q_{in}$,
indicating an assortative correlation. The same behavior is found
between non-reciprocal out-degrees (lower right plot of Fig.
\ref{knnnr}). Concerning the crossed correlations, we observe again
a positive correlation between the neighboring non-reciprocal
in-degree and the non-reciprocal out-degree but no noticeable
correlation in the opposite one, that is, the average non-reciprocal
out-degree of the reciprocal neighbors of a node is independent of
the non-reciprocal in-degree of that node (see lower left plot in
Fig. \ref{knnnr}). In summary, the analysis of the two-vertex degree
correlation behavior indicates that most of the structural
correlations of Web graphs are found in vertices connected by
reciprocal links. This type of links therefore represents an element
of particular interest in that they express the ordering principles
(beyond simple randomness) at the basis of the Web structure.

\subsection{The reciprocal subgraph}
\begin{figure}[h]
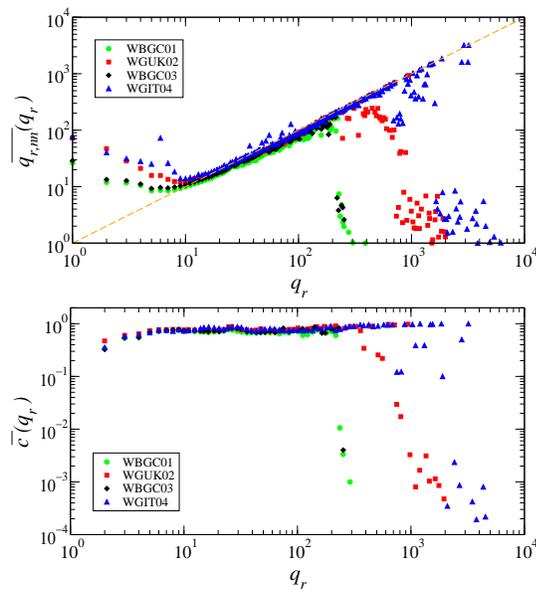

\centering $\begin{array} {c}
  \psfig{file=knnb2.eps,width=7cm} \\
  \psfig{file=ckb.eps,width=7cm}
\end{array}$
\caption{Average nearest neighbors degree (top) and degree-dependent
clustering coefficient (bottom) for the reciprocal links and for all
the samples.} \label{knnbckb}
\end{figure}

Very interesting information is provided by the study of how
reciprocal links are structurally organized among them. If we look
at the subgraph formed by the vertices and the reciprocal links we
can use the tools adopted for undirected graphs. A measure of the
two vertices correlation function is therefore expressed by
$\overline{q_{r,nn}}(q_r)$ (see Sec.~\ref{degree-degree}), i.e., the
standard measure of an undirected network if we identify reciprocal
links as undirected. As shown in Fig.~\ref{knnbckb}, this function
shows a first decrease, for $q_r<10$, followed by a linear increase
up to a critical value depending on the crawler. At high reciprocal
degrees, a cloud of points is populating the low r-degree region of
the average nearest neighbor reciprocal degree. This defines a
bi-modal pattern which indicates two different behaviors. The low
values cloud can be interpreted as a collection of star-like
structures, with central hubs connected to low degree nodes. This
effect is probably due to the ``home'' button in many Web pages that
belong to a bigger site. The linear behavior may have two different
interpretations. The first one is that the network is a tree in
which high degree nodes are connected to other high degree nodes.
The second one is that the network forms clique-like structures,
that is, groups of pages pointing simultaneously to each other.
To discern which scenario is more appropriate we inspect the local
connectivity properties of reciprocally linked vertices. Since we
can treat the reciprocal subgraph as an undirected one, we can probe
the local interconnectedness by analyzing the clustering coefficient
defined as the fraction of inter-connected neighbors of $j$:
$c_j=2\cdot n_{\tt link}/(q_{r,j}(q_{r,j}-1))$, where $n_{\tt link}$
is the number of reciprocal links between the $q_{r,j}$ reciprocal
neighbors of $j$. This quantity measures the density of
interconnected vertex triplets and it is therefore close to one in
the case of a fully interconnected neighborhood and zero in the case
of a tree structure. Global statistical information can be gathered
by inspecting the average clustering coefficient $\overline{c}(q_r)$
restricted to classes of vertices with reciprocal degree $q_r$. In
the first scenario, $\overline{c}(q_r)$ should be very small and
decreasing with the degree because of the tree-like structure. In
the second one $\overline{c}(q_r)$ should be significant and
independent of the degree. In Fig.~\ref{knnbckb} we show the
function $\overline{c}(q_r)$ which exhibits a high and constant
value followed by a cloud of points with very low clustering
coefficient at the same point where the function
$\overline{q}_{r,nn}(q_r)$ also splits. This indicates that the
organization of the reciprocal subgraph is a set of star-like
structures combined with cliques, or communities, of highly
interconnected pages. Very interestingly, this pictorial
characterization appears to be the same in all Web graphs
considered, pointing out to a genuine feature of the Web graph. The
present analysis identifies in the reciprocal subgraph an important
element that might help in decoding the structure of the WWW.
Finally we have to stress that the reciprocal component is surely
extremely important for the analysis and understanding of navigation
patterns and the network resilience to link removal.

\section{Outlook}

Contrary to what happened with the scrutiny of Internet maps, the
issue of sampling biases in the structure of the WWW has been left
almost untouched. The large size of the data sets has led to the
belief that the global properties were well defined in view of the
abundant statistics available. Noticeably, from the present
analysis, it appears that the resulting picture of the WWW structure
and its statistical characterization can be considerably affected by
the design of the tools we use to observe it. While some of the
basic properties are qualitatively preserved across different data
sets, other features and quantities are highly variable. This
results in a fuzzy picture of the WWW structure, where sampling
biases still play a major role. In other words, we are still in a
position where it is impossible to have a definite conceptual
framework to decode the structure of the global Web and how
effectively we can navigate, search, index, or mine the Web. The
present work thus highlights the need for a theoretical framework
able to approach a detailed analysis and understanding of the
sampling biases implicit in the most widely used crawling
strategies. In this sense, numerical studies of simulated
exploration of directed network models could be a starting point to
approach this problem and have a preliminary assessment of the
intrinsic biases induced by the crawling process. Finally, the
results presented in this paper are potentially helpful for
improving the design of future crawlers, not only regarding latent
biases. These applications are improved to a great extent when they
take advantage of the special hyperlink structure among web
documents and, at this respect, reciprocal links could play a key
role which has to be explored in more detail.

\section{ACKNOWLEDGMENTS}
We acknowledge the Stanford WebBase project and the LAW WebGraph
project for providing publicly available data. We would also like to
thank Filippo Menczer for helpful discussions and valuable comments.
This work is funded in part by the Spanish government's DGES Grant
No. FIS2004-05923-CO2-02 to M.~B., by a Volkswagen Foundation grant
to S.~F., by NSF award 0513650 to A.~V., and by the Indiana
University School of Informatics.


\appendix
\section{Degree-degree correlations:\\Quantitative definitions}
\label{APPA}We study the most significant two-point correlation
functions, taking into account that we can segregate the
neighborhood of each single node $i$ into neighboring nodes
connected to it by incoming links, the set $\nu_{in}(i)$, and
neighboring nodes connected to it by outgoing links, the set
$\nu_{out}(i)$. Following Eq.(\ref{eq_knn}), we can write
\begin{equation}
\begin{array}{lll}
  \overline{k_{in,nn}}(k_{in}) &=& \frac{1}{\kappa_{in,out}}\frac{1}{N_{k_{in}}}\sum_{i\in \Upsilon(k_{in})}
  \frac{\sum_{j \epsilon \nu_{in}(i)}k_{in,j}}{k_{in,i}} \\[0.2cm]
  \overline{k_{out,nn}}(k_{in}) &=& \frac{1}{\kappa_{out}}\frac{1}{N_{k_{in}}}\sum_{i \in \Upsilon(k_{in})}
  \frac{\sum_{j \epsilon \nu_{in}(i)}k_{out,j}}{k_{in,i}} \\[0.2cm]
  \overline{k_{in,nn}}(k_{out}) &=& \frac{1}{\kappa_{in}}\frac{1}{N_{k_{out}}}\sum_{i\in \Upsilon(k_{out})}
  \frac{\sum_{j \epsilon \nu_{out}(i)}k_{in,j}}{k_{out,i}} \\[0.2cm]
  \overline{k_{out,nn}}(k_{out}) &=& \frac{1}{\kappa_{in,out}}\frac{1}{N_{k_{out}}}\sum_{i\in \Upsilon(k_{out})}
  \frac{\sum_{j \epsilon \nu_{out}(i)}k_{out,j}}{k_{out,i}} .
\end{array} \label{eq_knn,in}
\end{equation}

These measures are normalized by the corresponding uncorrelated
values defined in section \ref{sec_degreedistributions} as the
heterogeneous parameters $\kappa_{in,out}$, $\kappa_{in}$, and
$\kappa_{out}$, in order to make them independent of the system size
and so comparable across samples.

The same quantities can be calculated when non-reciprocal and
reciprocal links are differentiated. Now, the neighborhood of each
single node $i$ is segregated into neighbors connected to it by
non-reciprocal incoming links, the set $\nu_{in}^{nr}(i)$, neighbors
connected to it by non-reciprocal outgoing links, the set
$\nu_{out}^{nr}(i)$, and neighbors connected to it by reciprocal
links, the set $\nu_{r}(i)$. The functions given in
Eq.~\ref{eq_knn,in} are valid whenever the in and out subscripts are
restricted to non-reciprocal links. When following only reciprocal
links, one can redefine them in a similar way:
\begin{equation}
\begin{array}{lll}
  \overline{q_{in,nn}}(q_{in}|r) &=& \frac{1}{\kappa_{r,in}}\frac{1}{N_{q_{in}}}\sum_{i\in \Upsilon(q_{in})}
  \frac{\sum_{j \epsilon \nu_{r}(i)}q_{in,j}}{q_{r,i}} \\[0.2cm]
  \overline{q_{out,nn}}(q_{in}|r) &=& \frac{1}{\kappa_{r,out}}\frac{1}{N_{q_{in}}}\sum_{i\in \Upsilon(q_{in})}
  \frac{\sum_{j \epsilon \nu_{r}(i)}q_{out,j}}{q_{r,i}} \\[0.2cm]
  \overline{q_{in,nn}}(q_{out}|r) &=& \frac{1}{\kappa_{r,in}}\frac{1}{N_{q_{out}}}\sum_{i\in \Upsilon(q_{out})}
  \frac{\sum_{j \epsilon \nu_{r}(i)}q_{in,j}}{q_{r,i}} \\[0.2cm]
  \overline{q_{out,nn}}(q_{out}|r) &=& \frac{1}{\kappa_{r,out}}\frac{1}{N_{q_{out}}}\sum_{i\in \Upsilon(q_{out})}
  \frac{\sum_{j \epsilon \nu_{r}(i)}q_{out,j}}{q_{r,i}} ,
\end{array} \label{eq_knn,inr}
\end{equation}
and the normalization terms in this case are
\begin{equation}
\begin{array}{lll}
  \kappa_{r,in} &=& \frac{\langle q_{r}q_{in}\rangle}{\langle q_{r}\rangle}\\[0.2cm]
  \kappa_{r,out} &=& \frac{\langle q_{r}q_{out}\rangle}{\langle
  q_{r}\rangle}.
\end{array} \label{kappa,dir}
\end{equation}

\balancecolumns 

\end{document}